# The role of spatial structures and social values in shaping local productive systems - New lessons from the wood-furniture cluster of Jepara, Indonesia

**Julien Birgi** [1]

*Sorbonne-Paris-Cité University, Paris, France*[1]
birgi.julien@gmail.com

**Abstract**

This paper revisits the well-known wood-furniture cluster of Jepara (Central Java, Indonesia) with new parameters inspired by the theories of industrial districts and clusters. So far, literature on this production centre, mostly focused on value chain analysis, has failed to explain how it could survive external shocks and rising pressure of global competition, without upgrading.

Based on a qualitative analysis encompassing the spatial organization and social structures supporting the productive system, this contribution of geography, sociology, and anthropology to development economics unveils a singular combination of ultraliberal business practices together with conservative social values, expressed in a unique arbour-like morphology. These qualitative results tend to show that, beyond the concept of flexible specialization identified by literature as the core driving force of the cluster, Jepara may be considered as a new avatar of Marshallian district in a context of subaltern globalization.

Such findings question the opportunities offered to subaltern towns in emerging economies enrolled in globalization to capitalize their own local social and environmental assets to develop labour-intensive activities capable of adapting an ever-changing deal imposed by external price-driven competition.

## 1. Introduction

Quite remote from Indonesia's main business hubs, the small town of Jepara and its eponymous regency of 1.2 million inhabitants are home to an impressive wood furniture cluster accommodating 10,000-15,000 firms that supply the global market, totalling 110,000-180,000 workers[1]. Such a concentration in a rather rural area has aroused curiosity of both Indonesian and foreign commentators since the late 1990s. But their focus on value chain analysis fails to explain how, in the long run, this atomized cluster could overcome external shocks, rising international competition, and shortage in wood resources, without upgrading. As a matter of fact, the bulk of literature has clung to the concept of flexible specialisation with little consideration to non-economic factors, no matter the key part they can play in shaping adaptative productive systems.

Indeed, economic activities cannot be considered as off-ground: they take place in a particular physical and social environment with whom they interact continuously. This notion of place, understood as the combination of interacting physical, social, and economic structures is core to local development theory, through the concepts of Industrial Districts (ID) (Beccatini, Bellandi et de Propris, 2009), clusters (Porter, 1998), positive feedback (Arthur, 1990), or territoire (Pecqueur & Zimmerman, 2004).

Overshadowed since the late 1980s by macroeconomics in a global context of triumphing liberalism and mass marketing, the empirical approach of local developmentalists attached to investigate the local context in which every productive system is distinctively rooted has recently regained attention in the academic and political discourse. Indeed, after the Asian Financial Crisis of 1997-98 already unveiled the fragility of outward development fostered by the New International Division of Labour (NIDL), the rising challenges of climate change, complexity of supply chains, geopolitical tensions, and pandemics are unveiling the hidden costs of economic globalization and arising interest in development rooted locally, particularly in secondary places remote from global cities enrolled in global competition, known in Southern Asia as subaltern globalization (Denis & Zerah, 2017).

---

[1] Assessing the size of this cluster, where informal businesses prevail, is challenging. Official statistics from the Central Bureau of Statistics (*Badan Pusan Statistik, BPS*) firm census (7,800 firms and 96,000 workers in 2018) clearly underestimate activity. Considering labour occupation statistics (with 303,657 workers in industry and crafts in 2019), more reliable than the earlier based on officially registered firm only, and the share of furniture in the regency industrial and crafts GDP (around 60%), we can extrapolate that total labour could top 180,000 in 2019. This figure is in line with literature estimates, that range between 16,300 firms and 75,600 workers (Roda et al., 2007) and 11,000 firms and 110,000 workers (Anggara et al., 2013).

## 2. Methods and Materials

In line with the empirical approach of the Italian school of ID, our research is based on a qualitative case study aimed at exploring the physical and immaterial structures and forces driving the cluster. Our initial concern was to identify a sample case not too big for such an in-depth analysis, but reflective enough of the entire productive system, so that we can generalize results. Based on the only comprehensive atlas of the cluster available (Roda et al., 2017), updated with detailed satellite image analysis revealing concentrations of workshops and factories, we selected the village of Tahunan. Located at the heart of the cluster and known as one of Jepara's most active furniture-making centres, this sample area of 35 ha accommodates the full range of value chain segments and morphologies found in the broader system.

We then carried out a systematic analysis of physical and social structures likely to drive development process locally. Personal background of furniture-makers, consistence of business communities, firm structure and managers' behaviour towards change were investigated through interviews with 148 stakeholders (including 31 company managers and entrepreneurs and 102 workers), selected according to purposive and maximum variation sampling methods (Palys, 2008). We also analysed the morphology of production facilities to identify where and how the segments of the value chain locate, grow, and interact with public space and residential functions, through a comprehensive inventory of 1,633 premises and morphological analysis of related private and public space. In total, we spent 160 days on the ground during 7 stays between October 2016 and May 2019. This empirical approach, starting from the field reality to explore past interactions between economic, social, and spatial entities at various scales (from the basic production unit to the local neighbourhood and the broader regency) is typical of French geographic systemic analysis.

The last step of our approach consisted of comparing our case results with the entire cluster. We used a large scope of data, including historical data from various Indonesian statistical bodies (especially the regional and local bureaus of statistics (BPS), a comprehensive review of available academic literature and grey sources (including official reports, press articles, planning documents and business seminar minutes), and interviews with 15 observers from business unions, local Government, and other key witnesses.

We will present hereafter our results with an overview of the historic background of the furniture industry in Jepara, a summary of value chain organization, and an attempt to characterize the entropic and development forces driving the productive system in the context of an everchanging global market. In a second part, discussion will lead us to confront these results with the main models of Marshallian agglomeration spillovers.

## 3. Results
### 3.1 Historic anchorage as a key factor of cluster emergence

Whereas literature about Jepara tends to focus on the most dramatic moments (whether booms or crisis) of the wood-furniture industry contemporary story, we believe that systemic analysis needs to consider the productive system in the long run.

The so-called Italian school has underlined the key role of history in forging productive systems based on the clustering of specialized firms, through the model of ID, defined as "a naturally and/or historically bounded place characterized by the presence and interpenetration of a community of people, and a production apparatus" (Becattini, Bellandi & De Propris, 2009, p.xvii). Indeed, long time is needed to build up trust, reputation, and sets of values that rule communities of entrepreneurs, just as it shapes step by step the place that accommodates their activities. Nonetheless, the abundant literature available about Jepara cluster does not tell much about its historic roots.

A popular myth reported by several of our respondents dates back the carving vocation of the town as far as the 13th century, when an outstanding artist was exiled there by the King of Mojopahit, turned jealous by the stunning beauty of the carving made after his wife. But we need to refer to historians of Southeast Asia (Reid, 1988) and Java (Lombard, 1990) to find early evidence of Jepara's economic role, witnessed by European merchants. In the 15th century, the city founded only one century ago, had already become a flourishing port on the route to Spice islands. That early insertion in long-haul maritime trade allegedly fostered shipbuilding and repair involving teakwood sourcing and carpentry skills. The fine wooden ornaments seen in Mantingan mosque testify the existence of a singular know-how that probably contributed to the city fame across Java. In the 19th Century, furniture-making was reported as a key industry, supplying the Javanese aristocracy and Dutch clients as far as Batavia and Holland (Andadari, 2008). Closer to our time, we could gather several testimonies confirming that Jepara re-emerged as a major furniture-making centre in the 1970s, thanks to prestige orders from the Indonesian Government, who also encouraged the creation of cooperatives of craftsmen, with the support of bilateral cooperation programs with European countries. However, it is only in the mid-1980s that exports started, when a handful of Western visitors impressed by the talent of local carvers introduced orders of Western-style ornamented furniture.

But perhaps more than tangible evidence from this past, the pride it arouses among local entrepreneurs regardless of their ethnic or personal background stresses the role of history in shaping a distinctive community of place as a condition for the cluster, a typical pattern of ID.

**3.2 A singular trajectory of development**

The contemporary development trajectory of furniture industry in Jepara is quite singular. Over the past 3 decades, exports have increased at an average yearly rate of 13%. But the growth curb is uneven, characterized by an early boom (+63% a year between 1991 and 1999), followed by a collapse (-63% between 2000 and 2001), and a steadier recovery (+5% yearly between 2001 and 2020), cadenced by 4–5-year economic cycles[2]. Obviously, shifts in trends reflect the ups and downs of national currency exchange, which has a direct impact on the international competitiveness of products whose cost of production is determined by local inputs (namely raw wood and labour, which account for 80-90%[3]). But beyond this key factor, the continuous inflating cost of ever-scarcer wood supplies and ever-growing pressure from foreign competitors (especially China and Vietnam) should have undermined the position of Jepara on the international market, all the more since it has also lost competitiveness nationally (the share of Jepara in Indonesian exports of furniture fell from 26% in 2000 down to 11% in 2020[4]) for not having improved quality.

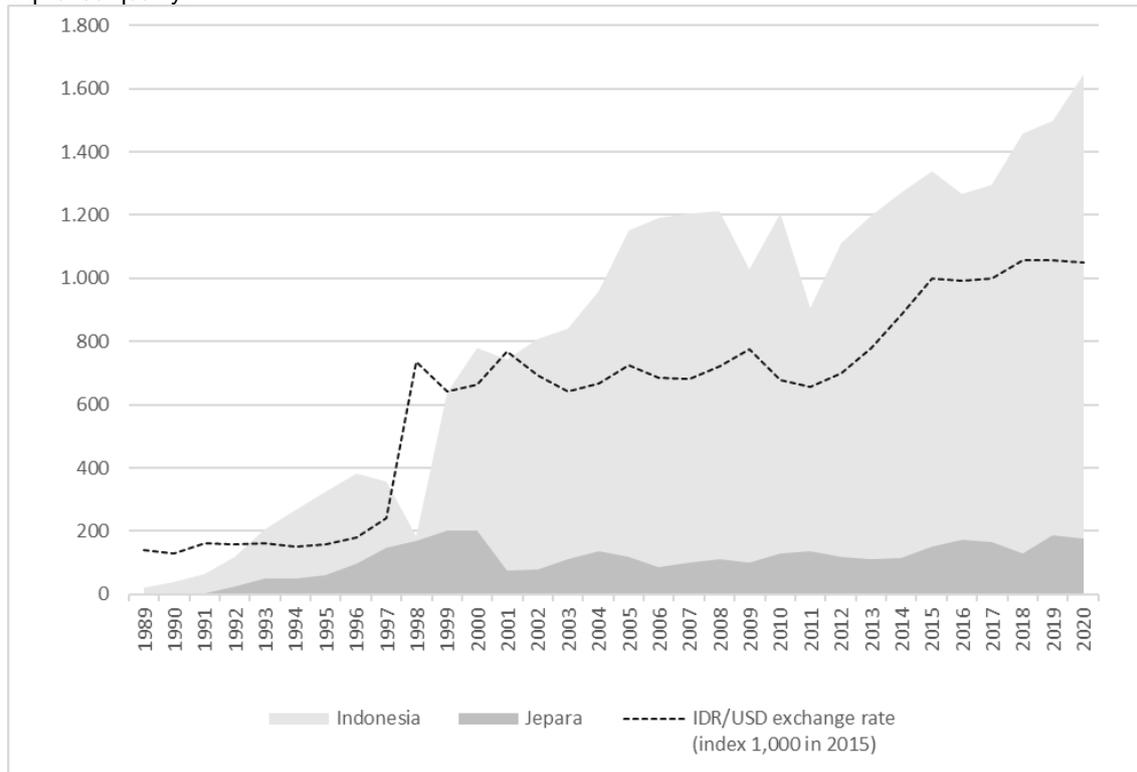

*Figure 1. Change in Jepara and Indonesian wood-furniture exports ($M.) with regards to currency exchange since 1989*

Yet, despite all those hostile forces, the furniture business keeps surviving: with regards to mainstream economic development models, such as the flying goose paradigm of Akematsu (1962) or the life cycle of Butler (1980), this record seems to contravene the Darwinian law of liberal economic leaving no alternative between upgrade or death. How can we explain it?

Literature about Jepara cluster has tried to explain periods of steep growth (Schiller & Schiller, 1997; Sandee et al., 2000) through the concept of "flexible specialization" (Piore et al., 1984). In a word, the furniture value chain is organized around the key figures of agents (*makelar*) who source producers for foreign and domestic buyers, entrepreneurs (*pengusaha*) who monitor production for them, and carvers and carpenters who work out the wood pieces either as subcontracting craftsmen (*pengrajin*) or contract workers (*borongan*). Upstream, log traders and sawmills dispatch raw wood, and downstream, shipping agents manage export formalities. The overwhelming majority of firms are household enterprises, employing informal labour and suppliers. *Pengusaha* hire cheap – not always skilled – labour while pengrajin use unpaid family members. These thousands of atomized players are interconnected by an infinite and highly volatile set of contracting ties, which include selling goods and services against money, and "free" of charge or charged work-sharing. These ties are characterized by their impermanence: business relationship is chiefly driven by day-to-day opportunities and continuously reconfigured[5]. Such an extreme flexibility optimizes almost instantaneously allocation of capital, labour, raw materials, and transportation means.

---

[2] After Jepara Central Bureau of Statistics (BPS)

[3] After BPS Central Java

[4] After BPS Jepara and UN Comtrade

[5] Our field survey has witnessed striking expressions of this flexibility. Cases of semi-finished products rerouted by over-bidders have been reported. Dereliction of duties by workers who found better temporary solutions are common. When getting orders too big for them, craftsmen tend to share or subcontract work at a cheaper price

As for recessions, literature points the lack of investment in quality of wood sourcing, training, machinery, design, and marketing, resulting in rather poor quality and low added value compared to international standards (Loebis & Schmitz, 2003; Andadari, 2008). This is also true for the 3 dozen big workshops and large factories[6], usually focused on assembling and finishing tasks, that cope with the ocean of plasma firms. These handicaps can be seen as the other side of the coin of flexible specialization. Indeed, this organization promotes price competition and reactivity at the expense of investment since there is no guarantee that in such a competitive context, capturing the returns is extremely challenging[7].

### 3.3 The role of a plastic physical environment in lubricating the cluster

Contemporary economists (including local development economists) tend to overlook physical space, owing to continuous progress in transportation of goods and people, and need for simplification of quantitative models. Largely absent from previous work on Jepara, we believe however that physical structures interact intensely with economic and social structures in either a developing or entropic manner.

In Jepara, the spatial layout of the wood-furniture cluster can be understood at three scales, namely the regency, the neighbourhood, and every single building premise. At the regency level, detailed satellite image analysis and ground visits confirm the findings of Roda et al. (op.cit.) who showed that firms and workshops aggregate within a radius of approximately 25 Km around Jepara town-centre, with density topping 200 firms/sq.km in the heart of Tahunan and decreasing in concentric circles to 15 firms/sq.km. within a radius of 10-12 Km, and 1-15 firms/sq.km. beyond. Getting out one's bearings in such a wide and scattered cluster is challenging. Hence, the relevance of areas specialized by product allowing ordering firms and buyers to know where to find them out. Local product specialization is not market-headed, however. Rather, it is the consequence of the principle of reproduction that rules furniture entrepreneurship in Jepara: relatives and neighbours tend to copy the successful business they can see next door. As a result, micro-clusters emerge around pioneering entrepreneurs working for a new market niche. Besides density and product variations, the value chain segments are also organized according to the road network. At the macro-scale, functions related to basic inputs (wood logs) and final outputs (showrooms and shipping agents) concentrate on Karangkebagusan Road (along the seashore) and Soekarno-Hatta Road (the main road connecting Jepara and Demak). This later supports heavy traffic, and as such offers high visibility for successful *pengusaha* willing to display their products. The first showrooms settled there in the late 1980s, but they really blossomed during the teak rush, when showy buildings were a popular sign of success aimed at both foreign clients and local fellows. Yet, the crisis of 2001-2002 and the boom of e-commerce since 2010 have deeply affected the commercial vocation of this road (32% of showrooms were derelict or vacant in 2015), while others had been converted into proximity retail outlets (Liantina, 2016), thus witnessing the capacity of real estate to meet fast-changing deal. Large assembling units prefer regency roads that offer a good compromise between easier truck manoeuvre and decent visibility of their output. Medium-sized workshops and sawmills opt for cheaper land on secondary streets still suitable for containers trucks. Household workshops are scattered in small streets and yards, the garage of the house being often used to store semi-finished parts. Carvers aggregate in the rear lanes that can be reached by motorbike. The road network thus plays a key part in the spatial organization of the value chain, shaping an original arbour-like structure where clusters of value chain segments stretch along a hierarchised grid of roads, streets, and paths. This grid determines land market prices[8], thus selecting segments of production according to their respective added value (which determines their capacity to pay for the cost of location).

But road status alone fails to explain the distribution of activities. At the neighbourhood level, we can note that local arrangements of value chain segments contradict the rationality of specialization, road access, and visibility. Such deviations from the rule can be explained by the pragmatic benefits drawn from using land inherited, bought, or rented from relatives. The founders of the clans who make out Tahunan community settled in the first half of the 20th century along the first roads opened in the area. Those primary houses being distant from each other, that scattered settlement allowed their descendants to build their own home and set up their furniture workshops around. Hence, a patchwork of living units made of clan-based clusters. Clan life is centred on proximity streets (*jalan lingkungan*) and family paths (*jalan keluarga*). These narrow roads suitable for pick-up vans suitable for local transportation of furniture pieces

---

implying loss in quality. Cheating on quality and payment is widespread, to such an extent that most interviewees say that they prefer not to work with family members because of higher risk to be stolen design or clients.

[6] After large and medium-sized firm directories issued by local Government and business unions, checked by satellite view detection and ground visits.

[7] For instance, Ebako furniture company, who run a factory in Semarang decided to put an end to their training program in Jepara as their subcontractors would use the new skills to work for competitors (interview with Mr. Hernawan, Executive Manager of Ebako, on 6 August 2016).

[8] In 2017-2018, they ranged between IDR 1.5-2.0 million/sqm. along Jalan Raya, 0.5 million along secondary roads suitable for container trucks, and 0.3 million along smaller streets (source: field survey).

and motorbikes are maintained and managed by the village local government (*Pemerintah Desa*), sometimes with a contribution of neighbours (either in land or in cash) based on the length of their property frontage and wealth. Street privatization for ceremonies is common, with self-made traffic signs to divert traffic and enrolment of village civil servants during weddings, funerals, or other communal ceremonies (*slametan*).

The narrow imbrication of residential, industrial, and commercial functions found at the neighbourhood scale is replicated in every building. After an initial training at a neighbour's or relative's where he can get skills, contacts, self-confidence, and savings, every *pengrajin* settles at home. Usually, he arranges a small shelter of 20-30 sqm. where he can carve freely, without interfering with domestic life. Facilities are expanded step by step, following demand, starting with workshops made of light-materials (bamboo poles, wood offcuts, and corrugated iron). As business grows, production is re-arranged. A *pengusaha*'s wife explains: "In 2000, our workshop was in front of the house. Then, when it was renovated in 2006, the workshop was moved besides the house and gave way to a showroom on the front part. And we also use a space in the backyard to dry wood and store and wood offcuts"[9]. Another example (illustrated by the sketch in appendix) shows how production tasks have sprawled into the original house, pushing away domestic spaces to new premises built at the ground floor (for key functions like kitchen and living room), or upstairs (for rooms and bathroom), no matter if it hides ornamented facades. Activity can overflow to the neighbours' garden and be re-arranged to draw the best from the plot lay-out, namely log storage, cutting and ovening in the backyard (served by a road), carving, assembling, and finishing in the middle, and packing and storage of final products in the front. If business keeps growing, the *pengrajin*, now become a *pengusaha*, may jump to a proper warehouse made of cement or bricks and tiles, plus a showroom, which is considered as an achievement of a successful business. During the furniture boom, premises would be preferably built on land already owned by the family, but when the crisis burst, many vacant premises became available, and rental spread. In local factories, we can note the same empirical flexibility in the use of both indoor and outdoor space, that contrasts with the rational lay-out of foreign-owned furniture-makers in a city like Semarang. Adjacent public open-air spaces – the roadside, or even entire family streets – are enrolled to increase the flexibility of private-owned assets. They can be used against financial compensation by a busy *pengusaha* to load/unload pick-up trucks, store semi-finished products during daytime, work out pieces of furniture, carry out restrictive tasks (like oiling or painting), park workers' motorbikes, as well as eating, resting, and chatting. These malleable spaces in the short-term differ from the long-term land reserves, where economic activity is banned. The so-called "empty land" (*tanah kosong*) found in the neighbourhood, usually natural plots covered with banana trees, can be used for free by neighbours for low-intensity domestic activities – with chickens wandering around for bugs, or waste disposal and burning – but not for money-earning activities. Though coexisting narrowly, residential and working areas are clearly separated[10], the richer the *pengusaha*, the more closed and restricted their house. Such a distinction can appear as a logical response to invasive activity, causing some nuisance (especially dust and noise) and ruled by liberal values in deep contrast with the conservativeness ruling the household and the neighbourhood as we are going to see.

### 3.4 Ultraliberal entrepreneurial values driving the cluster growth

Whereas it overlooked physical environmental factors, the theory of ID and clusters has underlined "the importance of appropriate contexts for sharing social experiences, moral values, and productive knowledge" (Becattini, Bellandi & De Propris, 2009, p.xvii), considered as expressions of Marshall's "industrial atmosphere".

In Jepara, the atomized cluster based on ultra-flexibility implies agile business communities and liberal entrepreneurial values. The first of them is "tolerant cosmopolitism": although native Jeparanese control the bulk of the industry[11], outsiders play a key part in production and trade. Massive rural migration from poorer regions of Central and East Java[12] has inflated labour and curbed wages, but it has also challenged local positions[13], and opened ties with other regions that make out potential market outlets. Though not as almighty as in other industries[14], Chinese Indonesians provide the cluster with financial and real estate services. Foreigners, who were the matchmakers between Jepara and the global market in the late 1980s, are still playing a key part locally. In such a small town, the swift and massive inflow of newcomers in a highly competitive economy, fuelling fierce labour competition and blatant inequalities, could have triggered serious social tensions. But evidence of conflicts is scarce, and outsiders seem to be well inserted locally (almost always through marriage with a native Jeparanese girl[15]), including those who adopt deviant attitude according to Indonesian societal and legal standards[16]. When times get harder, they return home or move to the big town to continue their business, thus creating new outlets for Jepara, where they keep sourcing skills and furniture pieces. No ethnic discrimination in business affairs (exclusively driven by pragmatism) is reported, a situation that differs much from guild-like craft centres, which are kept under tight control of exclusive social groups. Interestingly, all entrepreneurs, no matter their origin, refer to Jepara refer to the furniture-making founding myth, and claim the benefits of being players of a reputed place. Unlike guild-based clusters, where lineage excludes newcomers, the prevalence of *jus soli* over birth right makes Jepara an incubator of talents.

---

[9] Interview with Ibu Tuti, on 24 January 2018.

[10] Unlike other places in Java, where people can use their living room or some rooms to make furniture.

[11] Ante and post-teak rush literature figures are confirmed by our extensive field survey in Tahunan, where 93% of resident entrepreneurs surveyed are Javanese (80% of them being born in Jepara), 2% from Sumatra, and 5% from abroad.

[12] Between 1995 (just before the cluster take-off) and 2021, yearly demographic growth has been 1.89% in Jepara, above the provincial capital city of Semarang (1.60%) and provincial average (1.21%).

[13] It is interesting to note that most innovative *pengusaha* like that pioneer in e-commerce, the first printer of patterns on furniture, or that atypical manager attached to workers' empowerment are all outsiders.

[14] Schiller & Schiller (1997) argue that the lower profile of Chinese is due to the craft nature of the industry, low capital start-up requirements, key role of carpentry carving skills, employment flexibility, and intense daily communication required to deal with the Javanese pengrajin.

Besides cosmopolitism, other values allow to widen labour participation. Jepara may not be the stronghold of woman's power cultivated by the Indonesian *roman national* and local folklore[17], but spouses play an active part in the cluster and contribute to its flexibility (Andadari, 2008). They back their husband focused on production by handling administrative tasks, thus saving the hiring of an accountant. They can also run a side-business at home (small grocer's, plant nursery, or sewing service for instance) that brings a precious source of income when daily orders miss. Age discrimination is also less spread than in furniture factories[18], which results in the incorporation of the senior workforce in the productive system.

The plasticity of the Jepara class of entrepreneurs also applies vertically, since every incoming *pengrajin* can easily climb the social ladder to become a small *pengusaha*, or even a bigger factory-owner (and he can make the opposite way down just as easily if his business is doing bad). The cluster thus functions as a massive "social pump" for an ocean of unskilled labour, thus making an unlimited reservoir of cheap and reactive labour available for the global market.

Entrepreneurship is fostered by materialism, another key value more openly admitted here than elsewhere in Java, where money is (at least formally) often held in contempt. Jeparanese nouveau riche fancy showing off their success with posh houses, bling cars, and going out. Whether true or not, Indonesian outsiders and foreigners also often blame Jepara women for being venal (materialis), not-to-say women of easy virtue.

Individualism in economic affairs is striking. Families are cautiously kept away from household industry, as *pengusaha* fear they might be copied, cheated, or by-passed by relatives called in. Trust is not necessarily better with non-relatives, but at least problems can be sorted out more directly, or links cut off more easily than with relatives living close by. Thus, individualism clearly prevails over cooperation, in an ultra-flexible and liberal manner that guarantees reactivity and price competitiveness.

Fatalism or, better say, renouncement (*pasrah*), is a more surprising value dictating workers' but also entrepreneurs' behaviour, contradicting the principles of the Weberian "spirit of capitalism". Very few among the *pengusaha* polled envisage actions to adapt changes in demand and supply of inputs. Another example of *pasrah* is tokened by safety issues. Accidents are common and costly, even for independent workshops[19], but they are considered by Javanese victims and native managers as a demonstration of God's will that cannot be prevented and should be coped with[20]. *Pasrah* can be rightly seen as a major hurdle to production upgrade, at odds with typical business values of self-assertion and strategic vision. Yet, as underlined by Prof. Sudaryono from Gadjah Mada University in Jogjakarta[21], this well-anchored Javanese pattern may be considered as a response to external shocks that have affected overpopulated Java across centuries. Indeed, it would increase the resilience of workers and entrepreneurs, thus making them able to adjust market changes and economic shocks better than others, like Bapak Murdiono who accepts to "take little profit per sofa […]: as long as [he] can get orders, it should be fine"[22].

Even though our latest findings tend to show that some of the most distinctive patterns stressed above are fading away in an interesting move towards more stable business ties and attempts to differentiate through product personalization or proactive online marketing, they evidence the shift of Jepara from a reputed place of excellence to a centre for cheap and low-quality production, repositioning from *genius locus* rooted in unique artistic skills to competitive advantages based on more diffuse economic patterns of flexibility and resilience. Such an upheaval rises several questions about how to cope with extreme economic insecurity, save social cohesion despite fierce competition among individuals, and maintain the place identity and reputation challenged by the rampant irruption of global modernity.

---

[15] Mixed marriage is encouraged by the Indonesian Law, which bans ownership of physical assets for foreigners. It creates an unbalanced situation, where the wife owns land and real estate, and the husband banking accounts.

[16] Cases of heavy drinking, drug addiction, tobacco or drugs smuggling, and even children's prostitution are reported by both Indonesian and Westerners based in Semarang. But interestingly, none of these cases has been neither reported nor commented by Jepara people

[17] National Woman's Day, or Kartini Day (*Hari Kartini*), is celebrated every 21 April after the birthday of the young Jeparanese aristocrat who fought for Women's rights to education in the early 20th century.

[18] In factories, the "older generation" – beginning with people in their late 40s – tends to be excluded from factory work. As a result, those who cannot afford opening a shop (*warung*) or homestay (*kos*) must move back into farming or handyman (*serabutan*).

[19] Local social pressure requires to pay for medical expenses or give a financial compensation.

[20] All Western managers met stress their failure to enforce safety rules and blame their labour for lacking sense of personal responsibility.

[21] Discussion during a site visit, on October 2016.
[22] Interview on 18 February 2018.

### 3.5 Social conservatism as a social mattress to outweigh economic insecurity

When walking down the streets of Tahunan and talking with locals and outsiders, one can only be astonished by the deep contradictions between openness to the outside world and village-like mentality. Typical Javanese communal activities such as neighbourhood meals for family events (*slametan*), neighbourhood gatherings (*kumpulan rukun tetangga*), tontine (*arisan*), or collective prayers (*pengajian*) are as widespread as in the most rural parts of Java. Whereas business is done invidiously or at the nuclear family level, those communal practices lubricate daily interactions with the broader clan, which includes siblings, cousins, and close friends, thus acting as a counterweight to economic individualism.

Social activities do not only preserve the existing order. Solidarity, particularly strong inside the clan that shapes the Javanese society, plays a determining economic part in maintaining workforce available for unexpected demand. In between two jobs, contract workers tend to stay at home, waiting for a visit or a SMS to be called in. Such a behaviour could not be sustained if clan fellows did not share information and business opportunities, thus allowing *pengusaha* to meet any kind of order at any time. Social interactions among neighbours also allow widespread work-sharing, which allows to meet erratic demand by putting in common production capacities in an informal manner. At the nuclear level, the combination of social conservatism and entrepreneurship is embodied by marriage, which is almost always the starting point of setting up an independent business. Wedding brings assets, including financial capital (jewels), household goods (motorbike), real estate (family-owned land allowing to build a house that can later be used as a workshop), economic security (spouses tending to work), and wider business networks (brought by the cells of the parents-in-law).

The Jeparanese also care to maintain a strong link with their native place. He who left in search for a better life always comes back one day (usually to set up a furniture business). Even the most successful entrepreneurs who prefer to escape the burden of intrusive community life care to build a good house in their native village. A right of place applies, betrayed by local taxes charged by the village Head (*Kepala Desa*), supposedly for road maintenance, expected contribution to communal practices, and pressure to hire the village young idlers (known as *preman*, a term that refers to marginal people involved in delinquency) to upload containers and watch out the company assets at night. Local anchorage can be explained by objective assets offered by the place reputation: throughout Indonesia, being from Jepara means that one knows how to carve, and this reputation allows Jeparanese away from home to size all business opportunities related to furniture-making, no matter whether they are actually involved in this industry or not. Besides this powerful business card, the place offers concrete advantages, including a wide choice of possible suppliers close at hand, available premises in the bosom of family, and "free marketing" thanks to the mass of sellers clustered that attracts outside buyers.

Finally, the arbour pattern of the cluster has saved the environmental capital of the place. By channelling production and attached residential units along the road network (instead of spreading evenly over surrounding greenfield), industrial development has skirted rivers, woods, and cultivated fields, thus maintaining alternative resources for workers who can return to agriculture or during hard time. When compared with other forms of industrial development that irremediably convert natural land into industrial use, thus leading local people into a one-way development path, Jepara cluster appears as a resilient model that saves short-term and long-term economic alternatives.

### 3.6 The cluster benefits compared with development driven by industrial estates

We can wonder to what extent the very particular pattern of this productive system has improved standards of living in a developmentalist perspective. Macro indicators tend to show that the furniture industry has generated less income in cash in Jepara than in factory-driven centres like Semarang. Wages have increased more slowly than the regional average[23], and gaps in actual wages are even deeper[24], with unskilled contract workers (*borongan*) earning as less as Rp. 30,000-50,000 the days they get work[25]. Purchasing power has followed similar trends. At the peak of the furniture rush in 2002, Jepara ranked 4th among the 29 regencies in Central Java in terms of per capita non-food expenditure, but only 16th in 2015, with a rising share of food expenditure that now stands above regional average, thus reflecting local inflation in daily goods. Limited benefits generated by the cluster in terms of revenue should be outweighed by the possibility for workers to work at home, a condition praised by many Javanese who underline their attachment to spend time with their nuclear family, and a way to save. We also need to consider the benefits of saving transportation and lunch expenses, together with the possibility to carry out side-activities (such as gardening, petty trade, or riding a moto-taxi).

---

[23] The minimum legal wage in 2020 was only 75% that of Semarang, but it also lagged behind rural regencies less enrolled in industrialization.

[24] With official statistics assessing that Jepara salaried workers earn 56% of their Semarang counterparts, and more strikingly, only 72% of the provincial average.

[25] Irregular orders imply monthly income variations that can range between Rp. 400,000 and 1,200,000, compared to the fixed wage of Rp. 1,900,000 in a local garment factory.

Income stagnation in relative terms lies for sure with the inability of the cluster to upgrade from price-headed competition to more qualitative products, but also with a steep increase in labour supply. Both have challenged economic growth records, whose benefits are to be shared among more beneficiaries. In that sense, Jepara reminds us of the concept of involution introduced by Geertz (1963) for labour-intensive agriculture in Java. The undermining power of labour enrolment may be seen as a burden for the already employed locals, but it is an opportunity for outsiders, who are allowed to settle, gain experience, and climb the economic ladder. And the rural-urban continuum that prevails in dense Java allows to shrink labour supply in hard times, as jobless easily return to their home village[26]. Among them are skilled workers and talented entrepreneurs who spread across the country with the know-how and reputation gained in Jepara, who contribute to prevent the entropy of the system[27].

Even though the Regency Government has failed to pace up with the challenging population growth for public services[28], Jepara also embodies a sober mode of development based on self-construction and maintenance of production facilities, dwelling units, and neighbourhood roads, which has been able to industrialize free of investment in infrastructure. Only infrastructures directly related to the furniture industry, especially roads, have been improved in line with the pace of industrial development[29].

In our PhD dissertation (Birgi, 2022), we have also analysed after BPS statistics the impact of the furniture industry in Jepara compared to factory-based centres in the broader area of Semarang, and rural areas less touched by industrialization. We do not have time to present detailed results here, but in a word, our findings show that Jepara path of development has performed well in terms of Human Development Indicators, especially for the educational level of the young generation (with a 40-point flare of general education certificate graduates between 1985 and 2018, against 31-38 in Semarang region and 29 in the adjacent rural regency of Grobogan)[30]. The health situation is not as good however, as a consequence of overwhelming informal labour (implying that most workers cannot benefit from company health insurance cover generalized by the 2014 Governmental reform, resulting in 47% of Jepara regency citizens without health insurance in 2019 against less than 15% elsewhere) and struggling of local Government to provide basic sanitation and health services at the pace of demographic growth (16% of Jeparanese fell sick in 2018, against 11% in the city of Semarang and 13% in Grobogan, and 6% of Jeparanese children were born without medical assistance against 2% in Semarang and 0% in Grobogan).

Finally, the cluster cushions external shocks better than factory-based industrial centres. The share of population below the poverty line has increased less in Jepara than in the broader Semarang region after the Asian financial crisis of 1997-98 (respectively 1 and 5 to 10-point increase). The propensity of the system to "release" migrant labour and switch to alternative sources of income for the locals with more agility than places locked in industrial activity meets the needs for resilient economy in highly volatile markets.

## 4. Discussions: Jepara with regards to mainstream economic and spatial models

The pioneers of spatial economics Von Thünen (1826) and Alfred Marshall (1890) introduced theoretical models to explain how economic activities aggregate and generate increasing returns. The benefits of aggregation, known as Marshallian spillovers, include (1) a broader choice of materials, labour, and capital inputs, allowing the specialization of tasks, (2) easier access to outer markets thanks to business networks and fame of large production centres, and (3) emulation resulting from competition and cooperation in a business atmosphere. Marshallian spillovers are still considered today as the main driver of industrialization and resulting urbanization.

**4.1 Is Jepara a new avatar of industrial districts?**

In the 1970s, as main industrial regions based on Fordist-style production in America and Europe were challenged by the competition of low-wage countries imposed by the New international division of labour (NIDL), the performance of some Italian regions, characterized by a dense network of small and medium-sized firms operating in one single business sector, drew attention on alternative productive systems to big factories. Through empirical studies, Becattini (1979) and his followers from the so-called Italian School noted that economic growth in this "Third Italy" did neither rely on the Fordist model of vertical concentration prevailing in the historic industrial triangle Torino-Milano-Genoa, nor on massive public investments like in Mezzogiorno. That new avatar of Marshallian theory, which revisited spillovers from a networking perspective favoured by common values of entrepreneurship and "flexible specialization" (Piore & Sabel, 1984), opened new prospects for small towns in countries industrialized long ago.

---

[26] Between 2019 and 2020 for instance, as the covid-19 crisis was hitting the industry with a decrease of exports value by 5.2%, the regency population fell by a dramatic 5.8% (after BPS Jepara).

[27] See the case of Bukir furniture cluster in East Java, initiated by several entrepreneurs who had a record in Jepara (Mawardi, 2014).

[28] School/pop. and health centre/pop. ratios are respectively 17% and 47% lower than provincial average (BPS, 2015). This issue is well identified in Jepara planning documents28, that underline how demographic growth challenges public services, and therefore recommend to curb population inflow.

[29] The entire road network of the regency had been asphalted as soon as 2005, against 74% only in the province (BPS, 2015).

[30] Records for college are not as good (4.7-point increase in Jepara against 13.9 in the city of Semarang and 4.7-4.8 in its periurban regencies), but still better than rural areas (2.3 in Grobogan).

One the one hand, Jepara wood furniture cluster seems to meet the definition of Marshallian ID proposed by Becattini and al. (op.cit.)[31], namely "local specialisation"[32] including "horizontal (competitors), vertical (ordering firms and suppliers), and diagonal (related services and instruments) specialized activities". Being rooted in a position of interface of local resources and international demand since the 15th century, Jepara is also "incorporated into historically evolving conditions including technology, material needs, culture, policy and public administration, and characterized by a positive interaction of local forces with transglobal forces". "The economies 'external' to the firm but 'internal' to the district" are generated here by a common feeling of belonging to an outstanding place encouraging entrepreneurs to step in and attracting talents from outside, the kinetics of goods and information offered by the morphology of the arbour, and a development that saved pre-existing physical (arable land and woods) and social structures (the clan and its mechanisms of solidarity) that supports unlimited labour supply constantly available.

On the other hand, however, it contravenes to common rules identified by ID literature. Neither the "personal trust" nor the "decisive role in private and public investments to technical, human and relational capital played by locally-embedded centres of strategy and decision making" (Becattini et al., op. cit.) can be found in Jepara. Anti-developmentalist values – such as extreme individualism, short-term behaviours, and cheat – betray Jeparanese entrepreneurs' poor concern for their personal reputation. Their priority is to meet any order passing-by, which requires to be cheap and quick by reshuffling subcontracting cards, at the expense of trust. It is another type of "collective efficiency" (Piore and Sabel, op.cit.; Schmitz, 1999) here, driven by price and reactivity instead of quality and creativity as in ID.

**4.2 Jepara with regards to the cluster theory**

The concept of cluster appeared in the 1990s in the United States to designate "groups of companies and institutions co-located in a specific geographic region and linked by interdependencies in providing a related group of products and/or services" (Porter, 1998, p.78). Inspired by the case of Silicon Valley in California, it has gained unchallenged popularity among development economists, stakeholders, and policy-makers eager for replicating the recipe of that world class innovation pot (Vicente, 2018). Beyond their diversity underlined by prolific literature, the common feature of successful clusters lies with the narrow interactions among firms, research organizations, and funding bodies found in a single place. Although the common sense of the word could refer to a mere concentration of similar activities, literature underpins that economic clusters are innovation-headed and supported by intentional development policies aimed at promoting that "economy of knowledge" (Foray, 2009). Thus, the cluster model refers to "high-road" development strategies, where competitiveness is based on a continuous upgrade in quality, design, and innovation, which tends to discard labour or commodity-intensive industries ruled by price competition.

Our investigations in Jepara show different patterns. Unlike hi-tech clusters where Government, Universities, and research centres play a key role in designing and implementing complex long-run development strategies, public authorities are considered inefficient and corrupted by all local stakeholders. Their role regarding the local industry is limited to road provision and subsidies granted to very few to participate in business exhibitions. Other attempts, such as shared machinery, training, marketing centres, or publication of business directories and guides, have ended in failure. Several reasons can be accounted for that incompetence, including the informality of business-government ties that prevails over formalised expression of interests (like unions or chambers of commerce), the lack financial means and technical skills of local government theoretically empowered for local development affairs since decentralization, together with their negative perception of industrial growth, which is considered as a burden for the rising demand for public services caused by the inflow of labour rather than a benefit[33]. While vocational schools seem to play an active part in networking Indonesian furniture factory owners in Semarang (Birgi, 2022), it is not the case in Jepara where most staff – from carvers to managers

---

[31] The "canonical model of industrial districts [is defined by] a non-metropolitan context, formed by small towns; a set of values such as hard work, and a collective identity; and a social structure based on the preponderance of small entrepreneurs and workers [...]. Historical legacies such as extended family, sharecropping and peasant property, as well as local political subcultures [...] also influenced the genesis of 'diffuse industrialization' (ZEITLIN, 1992, p. 281).

[32] According to local BPS data, the furniture industry accounted in 2018 for 60% of large and medium-sized industries, 53% of exports, 28% of total workforce, and 15% of the GDP of the regency.

[33] Indeed, taxation rate in Indonesia (imposed on formerly registered firms only, who are a tiny minority in Jepara) is low (Alm, 2019), and only 20% of it is charged by local provincial and regency governments. The rare factories of our sample paying taxes on land assets and profit pay no more than 0.7% of their turnover. More firms contribute directly to the community, but those informal payments are symbolic (0.1% of their turnover on average).

– are self-taught. And, in a context of continuous re-arrangement of contracting ties, in which returns on investment are highly unreliable[34], research and innovation are embryonic.

Thus, far from the cluster model, the productive system appears there as a living but unconscious body that interacts continuously with an ever faster-changing outside world, juggling daily with limited local assets and hard external constraints in an unintentional but highly efficient manner, agile and blind like a shoal of fish.

**4.3 Can a low-road agglomeration of firms generate circular returns?**

Meanwhile the cluster model was disseminating, spatial economists of the so-called New Economic Geography focused on another type of benefit of firm concentration at the local (Arthur, op.cit.) or regional scale (Krugman, 1991), known as circular returns. Agglomerated firms indirectly fuel the local economy by consuming goods and services, and by paying wages, thus expanding the local pool of resources and clients, from which they can benefit themselves. According to Krugman and his followers, this virtuous snowball effect explains the increasing returns of metropolisation, understood as the joint growth of a local productive system together with the development of the local market.

Can this concept apply to non-urban areas like Jepara?

Blending available statistics and detailed data drawn from our sample of plasma businesses and firms, we assessed the spread-out of the furniture industry value chain. Though varying depending on the period and firm type, we found out that on average, raw materials accounted for roughly 50%, other material supplies for 10%, labour and self-employment for 25%, formal and informal taxes for 1% only, and profit for 20% of the selling price of the final product. Then, we assessed the share of this value remains locally. It is close to zero for wood supplies (which are sourced in other regencies), tops 90% for other material inputs (a direct effect of clustering of suppliers and subcontractors), equal to 100% for labour (thanks to the unlimited labour pool offered by the capacity of the productive system to incorporate new workers), and around 75% for taxes and profit (given the fluidity of the entrepreneurial ladder, which allows anyone to set up his own business). In total, we can estimate that out of a turnover of $100, $51 is captured by local stakeholders. This share is much higher compared with factory-based production centres like Semarang, where similar analysis led us to amounts ranging between $24 and $36 depending on their level of anchorage in their close environment (respectively large industrial estates and cottage factories). If we add the effect of workers' consumption as assessed by their household spendings structure and the level of residential services found in the area, the amounts reach $55, against $30 and $37 respectively[35].

Those results are based on too many assumptions and too small samples to be taken for granted in absolute terms. But, in relative terms, they tend to show that concentration of firms and workers generate Arthurian positive feedback for the local area, thus counterbalancing to some extent the low added value generated by a material and labour-intensive industry. In other words, clustering may appear as an alternative to high-road paths of development for regions deprived of the resources demanded to upgrade. Of course, price competition, exacerbated by the jungle liberalism that prevails in the cluster, maintains individual income for entrepreneurs and workers. Nonetheless, the wealth created by the industry allows to incorporate thousands of local and migrant unemployed or poor farmers, functioning as an inclusive poverty pump whose economic and social impact should not be overlooked[36].

Indeed, the cluster has focused on industry distributed on an original arbour pattern melting a hierarchised road network and clan-based capsules. Its malleability, with large areas of open-air appropriable space for production adjustments and alternative family farming, thus allowing furniture-making tasks to grow at home, swarm inside neighbourhood cells, and sprawl to the main roads to meet global buyers.

## 5. Conclusion

Through this qualitative exploration at eye level of the production system, our results offer an insight of the forces driving a low-road cluster widely described but poorly figured out by literature. They tend to show that a singular corpus of values and unique spatial organization at all scales have shaped generated a highly adaptative productive system able to sustain the constant price pressure of the global market, as well as the swift changes in the technological environment. Though unnatural to all appearances, the combination of liberal economic rules allowing production to adjust demand instantaneously and village-like social conservativeness providing social cohesion has allowed Jepara to maintain its position as an international production centre through thick and thin. This paper stresses the contribution of physical space to the cluster competitiveness. The arbour morphology optimizes land and infrastructure allocation (thus easing the blossoming and endogenous growth of businesses as well as the accommodation of migrant labour), lubricates the circulation of market information and goods that comes along with networking and binding of household firms, and offers free marketing through the window effect of showrooms. Its plasticity ensures that buildings and open-air spaces can adapt fast-changing demand for economic activities and residential needs of inflating labour, without upsetting the clan structure that sustains daily life, solves conflicts, provides solidarity when business fades away, and maintains spatial anchorage that contributes to maintain the fame of the production centres.

---

[34] For example, Andy Saidan, a Sumatra-born *pengusaha* who tries to create new designs by printing vintage patterns, regrets: "they copy my ideas so much that I can see my own products on sale off the road before I could even find a way to market them myself. But I can do nothing against this, it will happen anyway" (interview on 24 July 2017).

[35] The low multiplying factor of workers' consumption in Jepara is due to wages barely above the poverty line and induced share of self-consumption (including house building, food, and leisure), a situation similar to cottage factories workers but that differs from industrial estate workers where salaried work goes along formal money spending.

[36] Between 1985 and 2019, industrial labour in the regency has increased by 265,000 workers, most of them being migrants coming from the poorest countryside of Central and East Java.

Such virtues question the potential of this case to offer an alternative to well-known models of industrial development underpinned by an injunction to upgrade for regions of emerging countries remote from the main cities, in the context of labour-intensive industries. Indeed, this place could not boast the infrastructure, educative and research amenities, local mass market and cultural diversity and services of big cities attracting creative classes, capital-intensive investment, and higher added-value services according to neo-Marshallian literature of the NEG. By playing right the cards of its limited resources to lift hundreds of thousands of poor peasants out of poverty, Jepara case exemplifies a sober path of development on the international stage that saves local environmental and social capital. In that sense, it appears as a perfect incarnation of "glocalization", understood as a combination of genius locus and insertion in the global environment.

To what extent what we dare to consider as a relative success story is replicable? Though combined spontaneously in a singular and organic manner, the historic, environmental, and social conditions that determined the system viability can be found elsewhere in Indonesia and beyond. Many places of emerging countries enrolled in subaltern globalization can rely on a trading tradition rooted in history, densely populated hinterland, and intense community life able to overcome the swift changes induced by development. The question remains open, thus inviting to investigate further low-tech clusters in emergent Asia as a way to revisit Marshallian theory in the global Souths.

**Acknowledgement**

The Editorial Board would like to thank all the participants and other parties who have contributed in the journal.**References**

11contribution of physical space to the cluster competitiveness. The arbour morphology optimizes land and infrastructure allocation (thus easing the blossoming and endogenous growth of businesses as well as the accommodation of migrant labour), lubricates the circulation of market information and goods that comes along with networking and binding of household firms, and offers free marketing through the window effect of showrooms. Its plasticity ensures that buildings and open-air spaces can adapt fast-changing demand for economic activities and residential needs of inflating labour, without upsetting the clan structure that sustains daily life, solves conflicts, provides solidarity when business fades away, and maintains spatial anchorage that contributes to maintain the fame of the production centres.

Such virtues question the potential of this case to offer an alternative to well-known models of industrial development underpinned by an injunction to upgrade for regions of emerging countries remote from the main cities, in the context of labour-intensive industries. Indeed, this place could not boast the infrastructure, educative and research amenities, local mass market and cultural diversity and services of big cities attracting creative classes, capital-intensive investment, and higher added-value services according to neo-Marshallian literature of the NEG. By playing right the cards of its limited resources to lift hundreds of thousands of poor peasants out of poverty, Jepara case exemplifies a sober path of development on the international stage that saves local environmental and social capital. In that sense, it appears as a perfect incarnation of "glocalization", understood as a combination of genius locus and insertion in the global environment.

To what extent what we dare to consider as a relative success story is replicable? Though combined spontaneously in a singular and organic manner, the historic, environmental, and social conditions that determined the system viability can be found elsewhere in Indonesia and beyond. Many places of emerging countries enrolled in subaltern globalization can rely on a trading tradition rooted in history, densely populated hinterland, and intense community life able to overcome the swift changes induced by development. The question remains open, thus inviting to investigate further low-tech clusters in emergent Asia as a way to revisit Marshallian theory in the global Souths.

**Acknowledgement**

The Editorial Board would like to thank all the participants and other parties who have contributed in the journal.

**References**

Akematsu, Kaname. 1962. "A historical pattern of economic growth in developing countries". The Developing Economies. 1 (1), 3–25. DOI: 10.1111/j.1746-1049.1962.tb01020.x.
Alexander, Jennifer, and Paul Alexander. 2000. "From Kinship to Contract? Production Chains in the Javanese Woodworking Industries". *Human Organization* 59 (11), 106-116.
Alm, James. 2019. "Can Indonesia reform its tax system? Problems and options". Working papers 1906, Tulane University.
Andadari, Roos K. 2008. *Local Clusters in Global Value Chains. A Case Study of Wood Furniture Clusters in Central Java (Indonesia)*. Amsterdam: Thela Thesis.
Arthur, W. Brian. 1990. "Positive feedbacks in the economy". *Scientific American,* 262 (2), 92-99.
Becattini, Giacomo. 1979. "Dal settore industriale al distreto industriale. Alcune consideracion sull'unita di indagine dell economia industrial" [Sectors and/or Districts: some Remarks on the Conceptual Foundations of Industrial Economics]. *Rivista di Economia e Politica Industriale,* 1, 1-8.
Becattini, Giacomo, Marco Bellandi, and Lisa de Propris. 2009. *A Handbook of Industrial Districts.* Cheltenham and Northampton, MA: Edward Elgar.
Berry, Albert, Edgard Rodriguez, and Henry Sandee. 2002. "Firm and group dynamics in the Small and Medium Enterprise sector in Indonesia". *Small Business Economics,* 18 (1-3), 141-161.
Birgi, Julien. 2022. "Quand la forme des systèmes productifs industriels dans un pays émergent façonne les qualités du développement local. Le cas de l'industrie du meuble au nord de Java-Centre (Indonésie), de 1985 à nos jours". PhD diss, Institut National des Langues Orientales.
Clements, Corinna, Jeffery Alwang, and Ramadhani Achdiawan. 2019. "Value Chain Approaches in a Stagnant Industry: The Case of Furniture Production in Jepara, Indonesia". *Bulletin of Indonesian Economic Studies,* 55 (3), 341-365. DOI: 10.1080/00074918.2019.1576855.
De Rosnay, Joël. 1975. *Le Macroscope.* Paris: Seuil.
Denis, Eric, and Marie-Hélène Zerah. 2017. *Subaltern Urbanisation in India: An Introduction to the Dynamics of Ordinary Towns.* Berlin: Springer.
Ewasechko, Ann C. 2005. *Upgrading the Central Java Wood Furniture Industry: A Value Chain Approach.* Genève: International Labour Office.
Foray, Dominique. 2009. *L'économie de la connaissance* [The Economy of Knowledge]. Paris: La Découverte. DOI: 10.3917/dec.foray.2009.01
Geertz, Clifford. 1963. *Agricultural Involution: The Process of Ecological Change in Indonesia.* Berkeley CA: University of California Press.
Krugman, Paul. 1991. "Increasing returns and economic geography". *Journal of Political Economy,* 91 (3). 483-499.

Map 1: The furniture industry cluster in Jepara regency *(source: Author, 2020):*

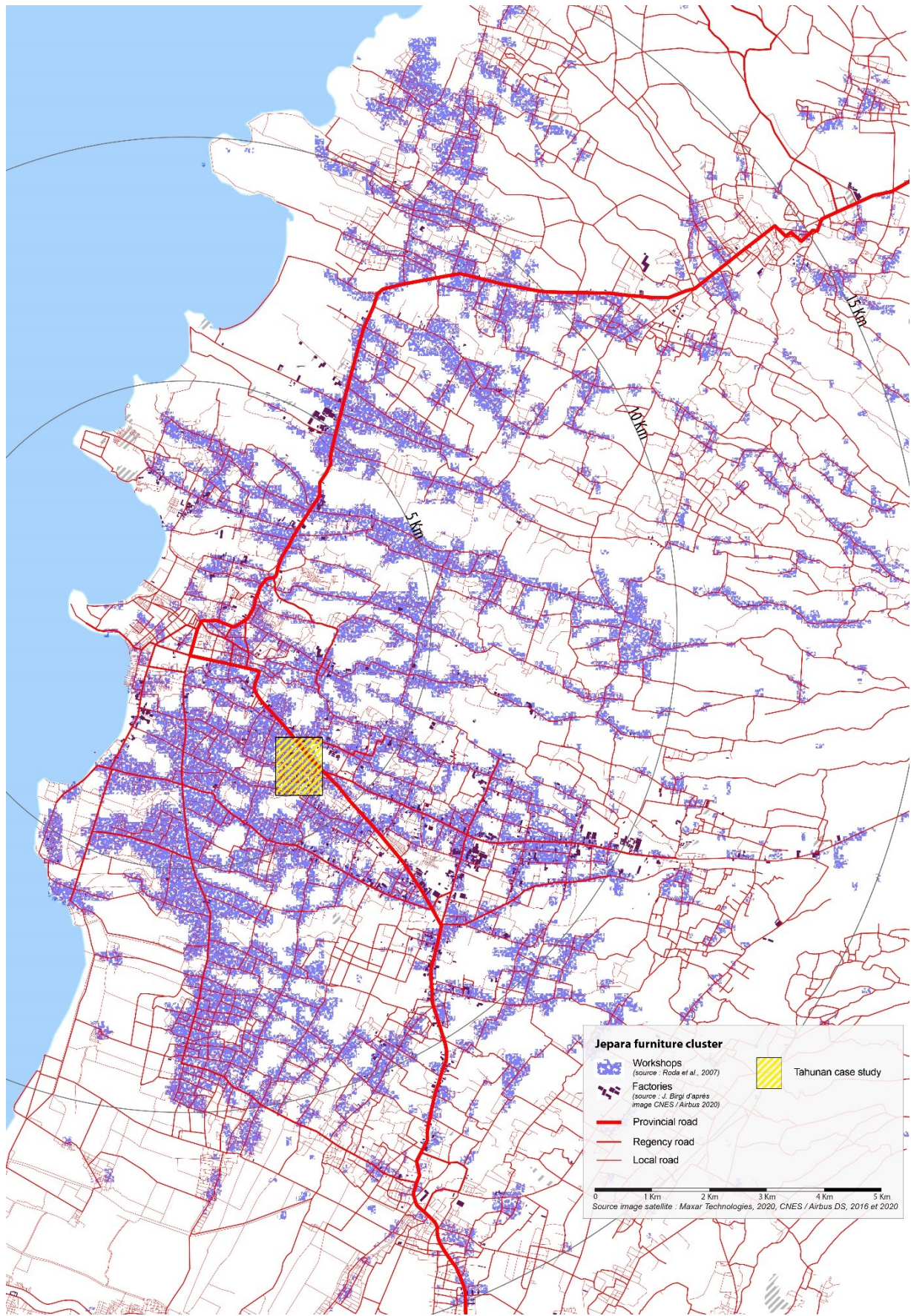

Map 2: Value chain segments distribution at the neighbourhood level, in Tahunan *(source: data and design: Author, 2020):*

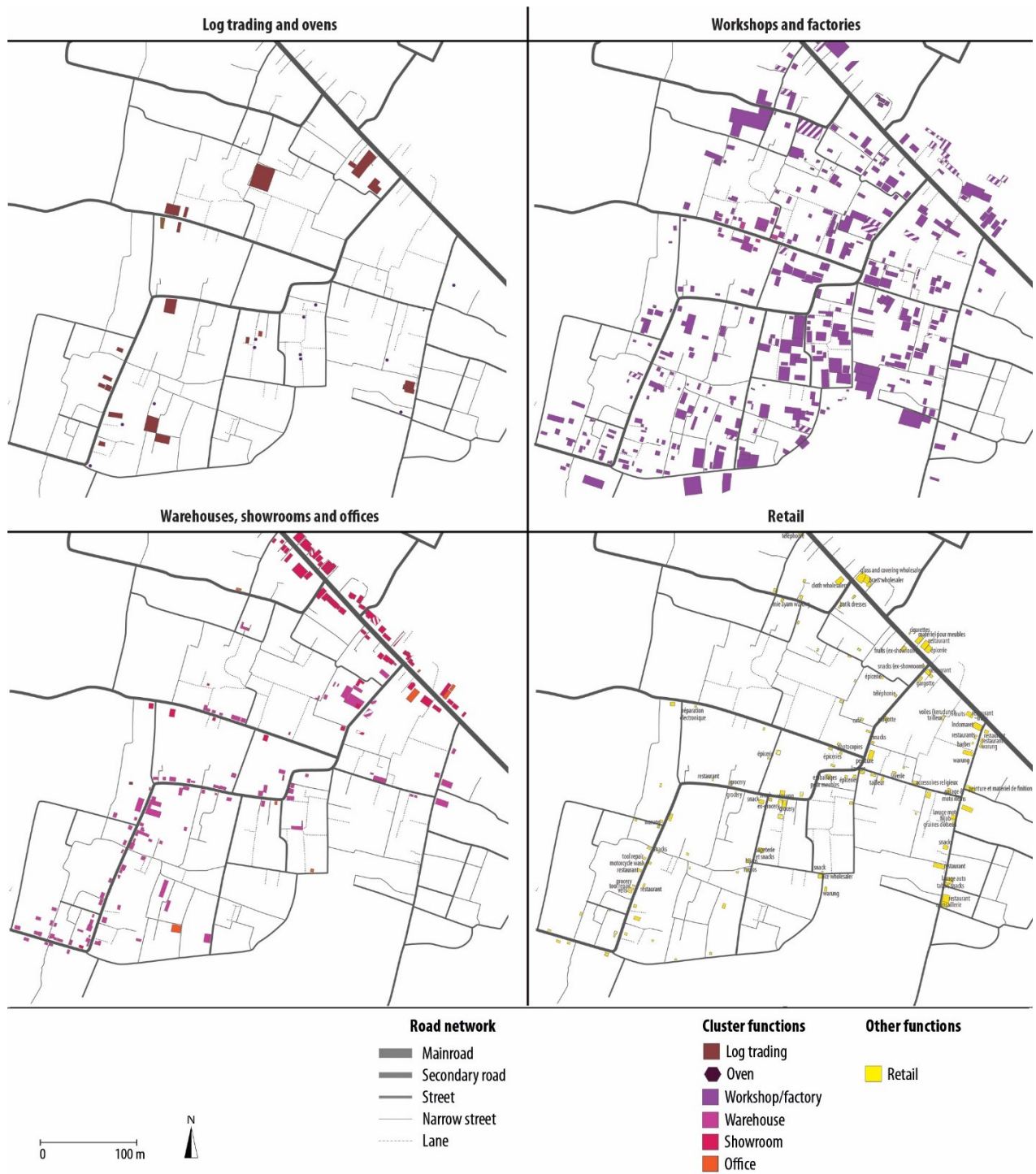

Fig 1: A typical case of flexible occupation of a pengusaha's house *(source: Author, 2020)*:

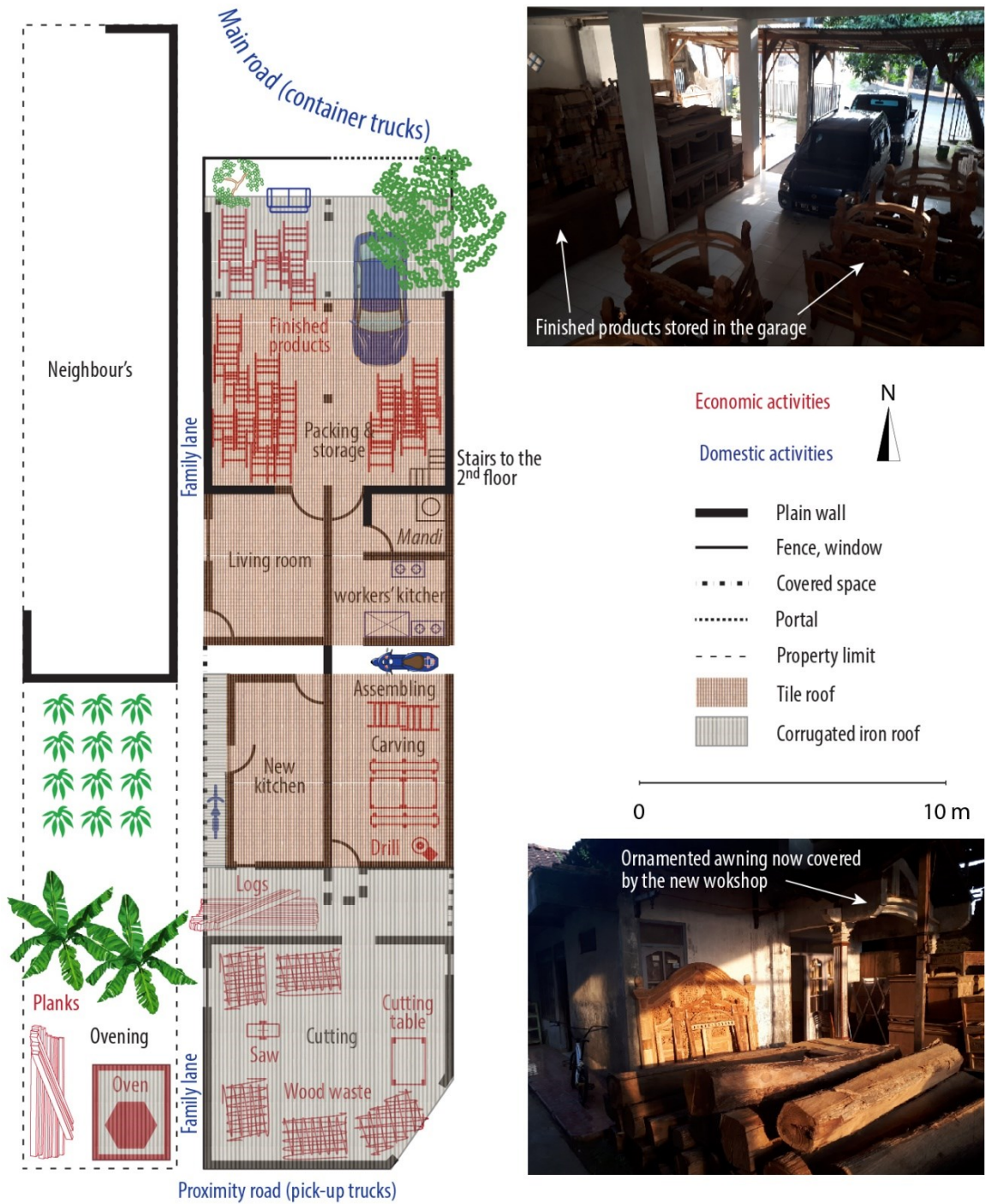